
\input harvmac

\newbox\grsign \setbox\grsign=\hbox{$>$}
\newdimen\grdimen \grdimen=\ht\grsign
\newbox\simlessbox \newbox\simgreatbox
\setbox\simgreatbox=\hbox{\raise.5ex\hbox{$>$}\llap
     {\lower.5ex\hbox{$\sim$}}}\ht1=\grdimen\dp1=0pt
\setbox\simlessbox=\hbox{\raise.5ex\hbox{$<$}\llap
     {\lower.5ex\hbox{$\sim$}}}\ht2=\grdimen\dp2=0pt
\def\simgreat{\mathrel{\copy\simgreatbox}}
\def\simless{\mathrel{\copy\simlessbox}}
\def\Hunit{\,{\rm km\,s^{-1}Mpc^{-1}}}
\def\GeV{\,{\rm GeV}}
\Title{HUTP-93/A020}{\vbox{\centerline{Could There Be Something}
\vskip2pt\centerline{Rather Than Nothing?}}}

\centerline{W. Daniel Garretson\footnote{$^1$}{Center for Astrophysics, Harvard
University, 60 Garden Street, Cambridge, MA \ 02138}}
\smallskip
\centerline{and}
\smallskip
\centerline{Eric D. Carlson\footnote{$^2$}{Lyman Laboratory of Physics,
Harvard University, Cambridge, MA 02138}}

\vskip .3in
\centerline{\bf Abstract}
\smallskip

\noindent There is increasing evidence that the universe may have a small
cosmological constant.  We suggest a scheme for naturally generating a
small cosmological constant.  Our idea
requires the presence of a discrete accidental symmetry which is
spontaneously broken by vacuum expectation values of the fields, and
explicitly broken by high dimensional operators in the Lagrangian.

\Date{07/93}

\newsec{Introduction}

Does the universe have an appreciable cosmological constant \ref\cpt{For
a recent review of the status of the cosmological constant, see S.M. Carroll,
W.H. Press, and E.L. Turner, {\sl Annu. Rev. Astron. Astrophys.}, {\bf 30},
499 (1992).}? 
The cosmological constant $\Lambda$ is defined by
\eqn\Ii{H^2 \equiv \left( {\dot a} \over a \right)^2 =
	{8\pi G \over 3} \rho_{\rm m} + {\Lambda \over 3} -
	{k \over a^2},}
where $H$ is the Hubble parameter, $a$ is the scale factor of the universe,
$\rho_{\rm m}$ is the energy density contained in the matter in the universe,
and $k = -1, 0$, or $+1$ depending on whether the universe is open, flat,
or closed, respectively.  It is useful to think of $\Lambda$ as an energy
density by defining
\eqn\Iii{\rho_\Lambda = {\Lambda \over 8\pi G},}
so that
\eqn\Iiii{H^2 = {8\pi G \over 3}\rho_{\rm tot} - {k \over a^2},}
where $\rho_{\rm tot} = \rho_{\rm m} + \rho_\Lambda$.

Since it is essentially impossible to measure $\rho_\Lambda$
directly, it is necessary to measure various observables in the
universe such as the Hubble parameter, the density parameter
$\Omega = 8\pi G\rho_{\rm tot}/3H^2$, and the deceleration parameter
$q_0 = - \ddot a a/\dot a^2$ .  Measurement of these
parameters can demonstrate the presence or absence of an appreciable
cosmological constant.

What, if anything, does particle physics predict for the cosmological
constant?  The naive value predicted by integrating the zero-point energies
of all possible modes in the vacuum with a cutoff at the Planck scale
disagrees with observations by some 120 orders of magnitude \ref\wein{
For a discussion of this point, see, for example, S. Weinberg, {\sl
Rev. Mod. Phys.}, {\bf 61}, 1 (1989), or, for a nontechnical discussion,
see L. Abbott, {\sl Sci. Am.}, {\bf 258(5)}, 106 (1988), or D.H. Freedman,
{\sl Discover}, {\bf 11(7)}, 46 (1990).}.  Hence, to a particle
physicist, the relevant question is not what the actual value of
$\Lambda$ is, but why it is so inconceivably small.  The fact that it
is so small has led many to conclude that it must be zero, and that
this value of zero must be enforced by some, as yet unknown, mechanism.

In addition to the standard assumption that $\Lambda$ is precisely
zero, it is common to assume that $\Omega$ is exactly equal to one.
This is due to the fact that, since $\Omega = 1$ is an unstable fixed
point in the evolution of the universe, it is very unlikely that it
would be close to one today unless it started out at a value that is
almost precisely one.  Since dynamical evidence indicates $\Omega \simgreat
0.2$, and the age of the universe implies $\Omega \simless$ 1--2, $\Omega$
must have been so close to unity in the early universe that many cosmologists
are convinced it must be precisely one.  In addition, if inflation
provides the correct explanation for the horizon and flatness problems
 \ref\guth{A.H. Guth, {\sl Phys. Rev. D}, {\bf 23}, 347
(1981).}, then $\Omega$ {\it must} be one today.

However, it is becoming increasingly apparent that one of these
assumptions ($\Lambda = 0$ and $\Omega = 1$) may have to be discarded.
There are several reasons for this, and we will briefly discuss the
most compelling of these, the so-called ``age problem''.  The age of the
universe today $t_0$ can be written in terms of the Hubble constant today
$H_0$ and the fraction of critical density $\Omega_i = 8\pi G \rho_i/3 H_0^2$
in the matter and vacuum as
\eqn\Iage{t_0 = H_0^{-1}\int_0^1 {dw \over \sqrt{1+\Omega_{\rm m}(w^{-1}-1)
         + \Omega_\Lambda (w^2-1)} }\; ,}
where $w=a/a_0$, the ratio of the size of the universe to its current size.
This formula assumes all the matter in $\rho_m$ is nonrelativistic; if some of
it is relativistic, the age will decrease somewhat.  It is easy to see that
the universe gets older as $\Omega_{\rm m}$ decreases or $\Omega_\Lambda$
increases.  For the case where $\Omega = \Omega_{\rm m} = 1$ (the ``standard''
assumption), we have the simple result $t_0={2\over3}H_0^{-1}$.  In figure
\fig\fone{$H_0t_0$ versus $\Omega_m$ for two cases: $\Omega=1$,
$\Omega_\Lambda=1-\Omega_m$ (long dashes); and $\Omega = \Omega_m$,
$\Omega_\Lambda=0$ (short dashes).  Note that, if $H_0 > 65\Hunit$ and
$t_0>12\,$Gyr, then the universe cannot have $\Omega=\Omega_m=1$,
$\Omega_\Lambda=0$.  Indeed, if $H_0>65\Hunit$ and $t_0>15\,$Gyr, then even
the values $\Omega = \Omega_m$, $\Omega_\Lambda=0$ aren't allowed.}
is plotted the age when we either drop the assumption $\Omega = 1$ while
retaining $\Lambda=0$ or drop the assumption $\Lambda=0$ while retaining
$\Omega=1$.
The point here is that,
if it can be shown conclusively that there exist objects in the universe
older than ${2 \over 3}H_0^{-1}$, then we will be forced to
conclude that, either $\Omega < 1$, or there exists some form of
energy density in the universe other than matter or radiation.  Indeed, if
the universe is older than $H_0^{-1}$, then we must include a cosmological
constant even if $\Omega_m < 1$.

This is significant, because it turns out that the best measurements
for $H_0$ seem to be clustering around values that are inconsistent
with the ages of the oldest stars in the galaxy assuming $\Omega =
1$ and $\Lambda = 0$.  The best fit ages of the oldest globular clusters
are $15-18$ Gyr, with lower limits in the range of $12-14$ Gyr \ref\stars{For
a reasonably concise introduction to this topic, see \cpt.}.  These conclusions
are extremely difficult to arrive at, and the final word may not be in, but
these results should be taken seriously.  Since the universe must be older
than the oldest stars, we will assume the universe must be older than 12 or
15 Gyr.

It is beyond the scope of this paper to undertake a general review of
the various methods of measuring $H_0$ \ref\jac{See, for example, G.H.
Jacoby, D. Branch, R. Ciardullo, R.L. Davies, W.E. Harris, M.J.
Pierce, C.H. Pritchet, J.L. Tonry, and D.L. Welch, {\sl PASP}, {\bf
104}, 599 (1992).}.  Roughly speaking, one can divide methods for
determining $H_0$ into two categories:  methods that measure
brightness, assume objects are standard candles, and determine the
distance by calibrating using nearby objects
whose distance can be found by other means; and methods that attempt
to measure distance directly using some physical phenomenon which, it
is hoped, is reasonably well understood.  The most reliable methods
for determining $H_0$ tend to fall into the first category, and those
discussed in \jac\ give values for $H_0$ of either $80\pm 11$ or $73\pm 11$
(in units of $\Hunit$), depending on how the distance to the Virgo
cluster is calculated.

The methods that attempt to measure distances directly tend to be less
well developed than the brightness methods, and thus are generally
thought to be less reliable at this time.  However, several of these
methods do tend to give lower values for $H_0$.  One example is the
Sunyaev-Zeldovich effect, which Birkinshaw, Hughes, and Arnaud
\ref\bha{M. Birkinshaw, J.P. Hughes, and K.A. Arnaud, {\sl Ap.J.} {\bf
379}, 466 (1991).} used to measure the distance to one cluster, from
which they found $H_0 = (40-50)\pm 12$ (the
range given is due to systematic uncertainties, while the error is due
mainly to random effects).  In addition, gravitational lens time
delays can, in principle, be measured to give the distance to lensing
galaxies, and this also has been applied to one lens system \ref\prh{For
the calculation of the time delay in this system, see W.H. Press, G.B.
Rybicki, and J.N. Hewitt, {\sl Ap. J.}, {\bf 385}, 416 (1992).} .  Again,
this gives a (slightly) low value for $H_0$ (the best value seems to
be $H_0 = 61\pm 7$, although this assumes
$\Omega = 1$).  However, there is a great deal of uncertainty in this
determination coming from uncertainties about the lensing mass
distribution, and uncertainties about the geometry of the universe
\ref\blanar{For a discussion of these points and a general review of
gravitational lensing, see R.D. Blandford and R.
Narayan, {\sl Annu. Rev. Astron. Astrophys.}, {\bf 30}, 311 (1992).}.

The one example of a method which attempts to measure distances
directly in which there are significant statistics and which is
thought to be fairly reliable at this time is the expanding
photosphere method, used on type II supernovae \ref\ske{For a description of
this method, see B.P. Schmidt, R.P. Kirshner, and R.G. Eastman, {\sl Ap. J.},
{\bf 395}, 366 (1992).}.  This method gives a value of $H_0 = 73 \pm 9$
\ref\sch{B.P. Schmidt, private communication.}, which is
consistent with the values derived from methods using brightness
determinations.

Although the results are inconclusive, it is fair to say that, with the
exception of a few techniques, measurements of $H_0$ seem to indicate
$H_0$ has a relatively high value (in the neighborhood of $70-80$).
Figure \fone\ shows, assuming $H_0 > 65 \Hunit$, that there is a conflict
between the age of the oldest stars and the assumptions $\Lambda=0$ and
$\Omega=1$.  In {\sl all} cases, $H_0t_0={2\over3}$ is
well outside the allowed ranges.  Interestingly, for the best fit
globular cluster ages, if $H_0$ lies in the given range, even the case
$\Omega = \Omega_m$, $\Omega_\Lambda = 0$ is ruled out, implying the
necessity that $\Omega_\Lambda \neq 0$.  It is only when one uses the
lowest allowed values for $t_0$ that one is not forced into a model
with $\Omega_\Lambda \neq 0$.  Thus the experimental case for nonzero
$\Omega_\Lambda$ is compelling, if not overwhelming.

In this paper, we examine the question of whether a small value for
the cosmological constant can arise, in any sense, naturally.  We find
that, making certain seemingly reasonable assumptions, it can.  In
section 2, we outline the basic scheme.  In sections 3 and 4, we
discuss some toy models in which this scheme is realized, and, in
section 5, we briefly discuss the issues of vacuum stability and
domain walls arising from this scheme.

\newsec{How a Small, Non-zero Cosmological Constant Might Arise}

Because there is no viable quantum theory of gravity, it is impossible to
predict the energy density of the vacuum.  Its measured smallness implies
that it may well be zero.  In this paper, we assume that the true minimum
of the particle potential will have exactly zero energy density.  Although
we do not promote a specific mechanism for achieving this, we note that
Coleman \ref\cole{S. Coleman, {\sl Nucl. Phys.} {\bf B310}, 643 (1988)} has
argued that, if wormholes exist, they would have the effect of
requiring the vacuum energy density of the ground state to be zero.
However, note that this is only a statement about the ground state.  There
may be other false vacua which would have non-zero energy density.  This is
explicitly stated in the work by Coleman \cole, but, indeed, if we are to
preserve the idea of inflation (a motivation for maintaining $\Omega=1$),
then we are forced to imagine false vacua with non-zero energy density.
Hence we assume that there is no mechanism for zeroing the energy density
in false vacua, only in the true vacuum.  We will not attempt to explain
the zero energy density of the true vacuum, but simply assume it.
In other words, we are not trying to understand why the universe should
have zero cosmological constant, but rather, given that it should have
zero cosmological constant, why it doesn't.

Our second assumption is that the universe is, in fact, in a false vacuum
state.  This false vacuum state must have an energy density very close to
the true vacuum energy.  Our goal is to explain the smallness of the
splitting.

In addition to the above assumptions, we also assume that the
full low-energy particle theory must include all
possible particle interactions; that is, all interactions that aren't
ruled out by any gauged or discrete symmetries of the full theory,
independent of whether or not they are renormalizable.
This is just an acknowledgment that we don't know what the full
particle theory should be, and that what we really have is a
low-energy effective theory valid up to some energy scale at which
we've integrated out all of the heavy physics.  Thus, when writing
down possible operators in the effective theory, we expect the
coefficients of the non-renormalizable operators to be suppressed
by the appropriate power of the mass scale at
which the effective theory becomes invalid ({\it e.g.}\ a dimension seven
operator would have a coefficient of the form ${\cal O}(1)/M^3$).

Why have we restricted ourselves to discrete and gauge symmetries?  Gauge
symmetries are known to exist, and it seems likely that they are preserved
by any non-renormalizable interactions.  Furthermore, discrete symmetries
can arise naturally as unbroken subgroups of spontaneously broken gauge
symmetries.  In contrast, it is believed that global symmetries must always
be broken by quantum gravity effects.  Hence gauge and discrete symmetries
may naturally be imposed on both renormalizable and non-renormalizable
interactions.

Given these assumptions, the scheme works as follows.  It is often the
case that, when one enforces a certain set of symmetries on operators
in a theory, certain subsets of the operators exhibit additional
``accidental'' symmetries.  It is our conjecture that the Lagrangian
of the universe will exhibit just such an
accidental symmetry---a symmetry which is only broken by a
fairly high dimensional operator.  This symmetry must be discrete.
Then, when the scalar fields in the theory acquire vacuum expectation
values (VEV's), this term will make different contributions to the vacuum
energy corresponding to transformations under the (broken) discrete symmetry.
If the operator
is sufficiently suppressed, the difference in vacuum energy coming
from the contribution in different vacua could provide the
cosmological constant.

\newsec{An Example Involving Discrete Symmetries}

An extremely simple example which illustrates the scheme we are
proposing is as follows.  Assume that we have a set of $N$ real scalar fields
$\phi_i$, $i = 1,\dots,N$, which transform under the permutation group
$S_N$, with the additional symmetry $\phi_i \to -\phi_i$, $\phi_j \to
-\phi_j$ for $i \neq j$.  The renormalizable terms in the scalar potential
consistent with these symmetries are
\eqn\IIIi{V(\phi_i) = -m^2\sum_i \phi_i^2 + \lambda_1 \left( \sum_i
\phi_i^2\right)^2 + \lambda_2 \sum_{i<j} (\phi_i^2-\phi_j^2)^2.}
Clearly, these terms also exhibit the symmetry $\phi_i \to -\phi_i$,
even though we didn't explicitly enforce this symmetry.
If we assume that all three of the parameters $m^2$, $\lambda_1$, and
$\lambda_2$ are positive, the minima of the potential will be given by
\eqn\IIIii{\langle\phi_i \rangle = \pm {m \over \sqrt{2N\lambda_1}} \; .}
These $2^N$ vacua are not all guaranteed to be truly degenerate.  They
break into two sets of $2^{N-1}$ vacua which are related among themselves
by the exact symmetry, and to each other by the accidental symmetry.  The
degeneracy between these two sets is broken by the first term one can write
down that breaks this accidental symmetry, which is the term
\eqn\IIIiii{{\lambda_0 \over M^{N-4}}\phi_1\phi_2\cdots\phi_N\; .}
Thus, in this model, there exist two discrete sets of minima where the
difference in the vacuum energy between the two sets of minima is
given, at lowest order in the VEV's, by the term proportional to \IIIiii.

It is easy to calculate how large $N$ has to be in order to get
specific values for the cosmological constant.  If the $\phi_i$ get a
VEV that is of order the weak scale ($\sim 100$ GeV), then this term
would make a contribution of $\sim 10^8(100\GeV/M)^{N-4}\GeV^4$ to the
vacuum energy density.  Thus, if we want a cosmological
constant with $\rho_{\rm vac} \sim 10^{-46}\GeV^4$, and if we assume
that the heavy mass scale is the Planck mass ($\sim 10^{19}\GeV$),
then we require $N = 7$.  Alternatively, if the $\phi_i$ get a VEV of
$\sim 10^{16}\GeV$ (that is, something like the GUT scale), then, again
using the Planck mass as our heavy mass scale, we require $N = 40$.


\newsec{An ``Accidental'' Symmetry in ${\rm SO}(N)$}

An alternative approach to using discrete symmetries would be to use gauge
symmetries, for example, ${\rm SO}(N)$.
For this example, we have again chosen a relatively simple gauge group
where one can, in some sense, intuitively understand the accidental
symmetry and how it arises.  For the moment, let us restrict ourselves
to a theory containing an arbitrary number of scalar fields $\Psi^i$ that
transform under the vector representation of ${\rm SO}(N)$.  If we label
the components of $\Psi^i$ by $\psi^i_\mu$, then, in general, any
potential one could write down would have an accidental symmetry as we
have described under which $\psi^i_1 \to -\psi^i_1$ for all $i$
(more generally, any odd number of components of the $\Psi^i$ can change
sign under this symmetry).  It should be apparent that this symmetry
(which we will call ``parity'') is not, in fact, a symmetry of the
theory since the determinant of the operator that gives this
transformation is equal to $-1$.

How does this parity symmetry arise?  Recall that, in the tensor
representations of ${\rm SO}(N)$, the only invariant tensors are the delta
symbol $\delta_{\mu\nu}$, and the fully anti-symmetric epsilon
symbol $\epsilon_{\mu_1\mu_2\cdots\mu_N}$, along with any other
tensors formed by multiplying these tensors together in various
combinations.  Because the $\delta$ tensor is invariant under parity while
the $\epsilon$ tensor changes sign, only terms involving the $\epsilon$
tensor can be parity noninvariant.  The smallest such term will be one in
which $\epsilon$ is contracted with $N$ distinct $\Psi^i$'s.  Note that,
in order for this term to actually break the degeneracy of the vacua, we
require that there be at least $N$ $\Psi^i$'s in the theory, and that their
VEV's span the ${\rm SO}(N)$ space.


To be more specific, let us imagine an ${\rm SO}(2n)$ symmetric theory with
one scalar field $\Phi$ in the adjoint (antisymmetric tensor) representation
with components $\phi_{\mu\nu}$.  The renormalizable portion of the potential
is given by
\def\nt#1{{1\over 2n}{\rm Tr}(\Phi^#1)}
\eqn\eIVi{V(\Phi) = m^2 \nt2 + \lambda_1 \left( \nt2 \right)^2 + \lambda_2
\left[ \nt4-\left(\nt2\right)^2 \right] \; ,}
where we have used a matrix notation for $\Phi$.  If $m^2$, $\lambda_1$,
and $\lambda_2$ are all positive, the minimum of the potential will be
given by
\eqn\IVii{\langle\phi^0_{\mu\nu}\rangle =
	{m \over \sqrt{2\lambda_1}}\pmatrix{0 & \mp 1 & 0 & 0 & & 0 & 0 \cr
	\pm 1 & 0 & 0 & 0 & \cdots & 0 & 0 \cr
	0 & 0 & 0 & - 1 & \cdots & 0 & 0 \cr
	0 & 0 &  1 & 0 & & 0 & 0 \cr
	& \vdots & \vdots & & \ddots & \vdots & \vdots \cr
	0 & 0 & 0 & 0 & \cdots & 0 & - 1 \cr
	0 & 0 & 0 & 0 & \cdots & 1 & 0 \cr}.}
where $\mp$ and $\pm$ occur in only the first row and column.  These minima are
related by the parity symmetry and degenerate to this level.

Using this, let us now look at the parity violating contributions to
the scalar potential (the extension of our definition of parity to
tensor reps of ${\rm SO}(2n)$ is simply that any component
$\phi_{\mu_1\mu_2\cdots\mu_m}$ where an odd number of the $\mu_i$
equal 1 changes sign under parity).  The lowest
dimensional operator that we can write using $\Phi$ that violates
parity has the form
\eqn\IViv{V_{\rm break} = {\lambda_0 \over n!\,M^{n-4}}
        \epsilon_{\mu_1\cdots\mu_{2n}}
	\phi_{\mu_1\mu_2}\phi_{\mu_3\mu_4}\cdots\phi_{\mu_{2n-1}\mu_{2n}} \; .}
Assuming, once again, that the vacuum expectation value is of order the
weak scale, and the large mass is of order the Planck scale, we can get a
cosmologically interesting vacuum energy density for $n=7$.

\newsec{Domain Walls and Vacuum Stability}

With this scheme in mind, one might wonder whether we must, of
necessity, have domain walls separating regions in the true vacuum
from regions in the false vacuum.  Similarly, one might wonder whether
high energy interactions in a region that is in the false vacuum might
cause bubbles of true vacuum to nucleate, thus erasing any
cosmological constant.

The first problem is easily solved if the universe undergoes a period
of inflation after the universe ``chooses'' its vacuum state.  The
inflationary epoch then inflates the domain walls out beyond the
observable horizon, and the problem is solved in the same way that
similar problems with, for example, monopoles are solved \ref\Sato{K. Sato,
{\sl Phys. Lett.}, {\bf 99B}, 66 (1981).}.

As for the question of vacuum stability, it is easy to see that the
false vacuum will, in fact, be stable.  Let us assume that the
universe is currently in the false vacuum, and calculate the critical
radius at which a bubble of the true vacuum would expand and fill the
universe.  We do this by considering the energy associated with a
bubble of true vacuum of a given radius $R$.  This is given by
\eqn\Vi{E_{\rm bubble} = -{4 \over 3} \pi R^3 \rho_\Lambda +
	4\pi R^2 \sigma,}
where $\rho_\Lambda$ is the energy difference between the true vacuum and
the false vacuum and $\sigma$ is the surface tension, typically of order
$v^3$, where $v$ is the VEV which breaks the accidental symmetry.  If
$dE_{\rm bubble}/dR$ is negative, the bubble will grow; for smaller bubbles,
it will shrink.  Setting $dE_{\rm bubble}/dR=0$ we find that the critical
bubble size is given by
\eqn\Vii{R_{\rm crit} = {2\sigma \over \rho_0} = \left(75 \Hunit \over H_0
\right)^2 \Omega_\Lambda^{-1} \left( \sigma \over (100\GeV)^3\right)\;
10^{21}\,{\rm Lt-yr} \;.}
In other words, if the scale of the VEV separating the true vacuum
from the false vacuum is around the weak scale, then, for a bubble of
the true vacuum to nucleate and grow,
it would have to form with a size that is at least $10^{10}$
times the size of the observable universe!  Clearly, the possibility
of bubble nucleation is not a problem for the cosmological constant in
this scheme ({\it i.e.}\ the difference between the false and true vacua is
irrelevant as far as particle physics is concerned).

\newsec{Conclusions}

A small cosmological constant is not as difficult to generate as might be
imagined.  The appearance of a spontaneously broken accidental discrete
symmetry, which is broken by non-renormalizable terms, naturally leads to a
small splitting between two nearly degenerate vacua.  These accidental
symmetries may occur in these models due to imposed symmetries, either
discrete or gauged.

If the lower of the two minima of the potential has a naturally zero
cosmological constant, then a universe stuck in the other minimum will
have a very small apparent cosmological constant.  Such a constant may help
resolve the universe's ``age problem.''  To avoid a universe where nearby
regions are in distinct minima, implying the existence of domain walls, it
is necessary to assume inflation occurred at or after the time of spontaneous
symmetry breaking.  Because of the smallness of the splitting between the true
and false vacua, there is no danger of the universe preferring one vacuum to
the other, nor is there any risk of tunneling to the true vacuum from the
false vacuum.

Unfortunately, the specific models we have proposed are unconstrained,
since the scale of the symmetry breaking and the scale of the
non-renormalizable terms, are completely unconnected to anything we know.
A natural scale for the appearance of non-renormalizable terms is the
Planck scale, but the scale of symmetry breaking is still free.  We view
this as an unattractive feature of our models.  This problem can be removed
in both of the models proposed simply by promoting the fields to ${\rm
SU}(2) \times {\rm U}(1)$ doublets with appropriate hypercharge so that
they can acquire a neutral VEV.  Then the scale for this VEV can be shown
to be at or slightly below the weak scale.  Unfortunately, this introduces
a huge number of additional doublets into the standard model.  This may
cause problems with oblique corrections to the
standard model.  Preliminary estimates indicate that the $S_N$ model (which
introduces only 7 new doublets for $N=7$) is probably allowed, but the
${\rm SO}(14)$ model (with 91 doublets) is in trouble.  Such a theory would be
more testable than the one proposed here, but would still be little more
than a toy model.

Much more promising is the idea of having the symmetry group we impose be a
grand unified theory (GUT) symmetry.  Several complications ensue, such as
the necessity of using a high power of $M_{\rm GUT}/M_P$ to get an
interesting cosmological constant.  But we still feel this is the most
promising direction for further development.

\newsec{Acknowledgements}

We would like to thank S.~Coleman, B.P.~Schmidt, and S.~Glashow for helpful
conversations.  This research was supported by the National Science Foundation
under contract NSF-PHY-92-18167 and by the Texas National Research
Laboratory Commission under grant FY9206.  D.~Garretson is supported
under a National Science Foundation Graduate Research Fellowship.

\listrefs
\listfigs
\bye